\documentclass[a4paper,11pt]{article}
\usepackage{pos}

\title{Long-distance hybrid spin-dependent and hybrid-quarkonium
mixing potentials from lattice gauge theory}
\ShortTitle{Hybrid spin-dependent and hybrid-quarkonium
mixing potentials}

\author*[a]{Vilija de Jonge}
\author[b]{Paul P\"utz}
\author[b]{Antonio Vairo}
\author[a,c]{Marc Wagner}

\affiliation[a]{%
  Institut f\"ur Theoretische Physik,
  Goethe-Universit\"at Frankfurt am Main, Max-von-Laue-Stra{\ss}e 1, D-60438 Frankfurt am Main, Germany}
\affiliation[b]{%
Technical University of Munich, TUM School of Natural Sciences, Physics Department,
James-Franck-Str.\ 1, 85748 Garching, Germany}
\affiliation[c]{%
Helmholtz Research Academy Hesse for FAIR, Campus Riedberg, Max-von-Laue-Stra{\ss}e 12,
D-60438 Frankfurt am Main, Germany}

\emailAdd{dejonge@itp.uni-frankfurt.de}
\emailAdd{paul.puetz@tum.de}
\emailAdd{antonio.vairo@mytum.de}
\emailAdd{mwagner@itp.uni-frankfurt.de}

\abstract{We present $SU(3)$ lattice gauge theory results for the $\mathcal{O}(1/m_Q)$ hybrid spin-dependent and hybrid-quarkonium mixing potentials. The only existing lattice computation of these potentials is limited to quark-antiquark separations smaller than approximately $0.5\,\text{fm}$. 
In this work, we extend the lattice results up to around $1.25 \, \text{fm}$ using large-volume, high-statistics simulations at several lattice spacings and gradient flow times, providing the first lattice data in this previously unexplored regime.}

\FullConference{The 43rd International Symposium on Lattice Field Theory (Lattice 2026)\\
July 26 to August 1, 2026\\
University of Maryland, College Park, USA\\}

\begin{document}
\maketitle

\section{Introduction}
\label{sec:introduction}

Heavy hybrid mesons consist of a quark-antiquark pair $\bar Q Q$ (with $Q \in \{ c, b \}$) and a gluonic excitation. The latter contributes to the $J^{PC}$ quantum numbers of the system and distinguishes a hybrid meson from conventional quarkonium states, which also consist of a $\bar Q Q$ pair, but without excited gluons. Conventional and hybrid quarkonia can share the same $J^{PC}$ quantum numbers and therefore mix, i.e.\ physical quarkonium states can be superpositions of conventional and hybrid quarkonium components.

\begin{figure}[htb]
    \centering
    \includegraphics[width=0.5\linewidth]{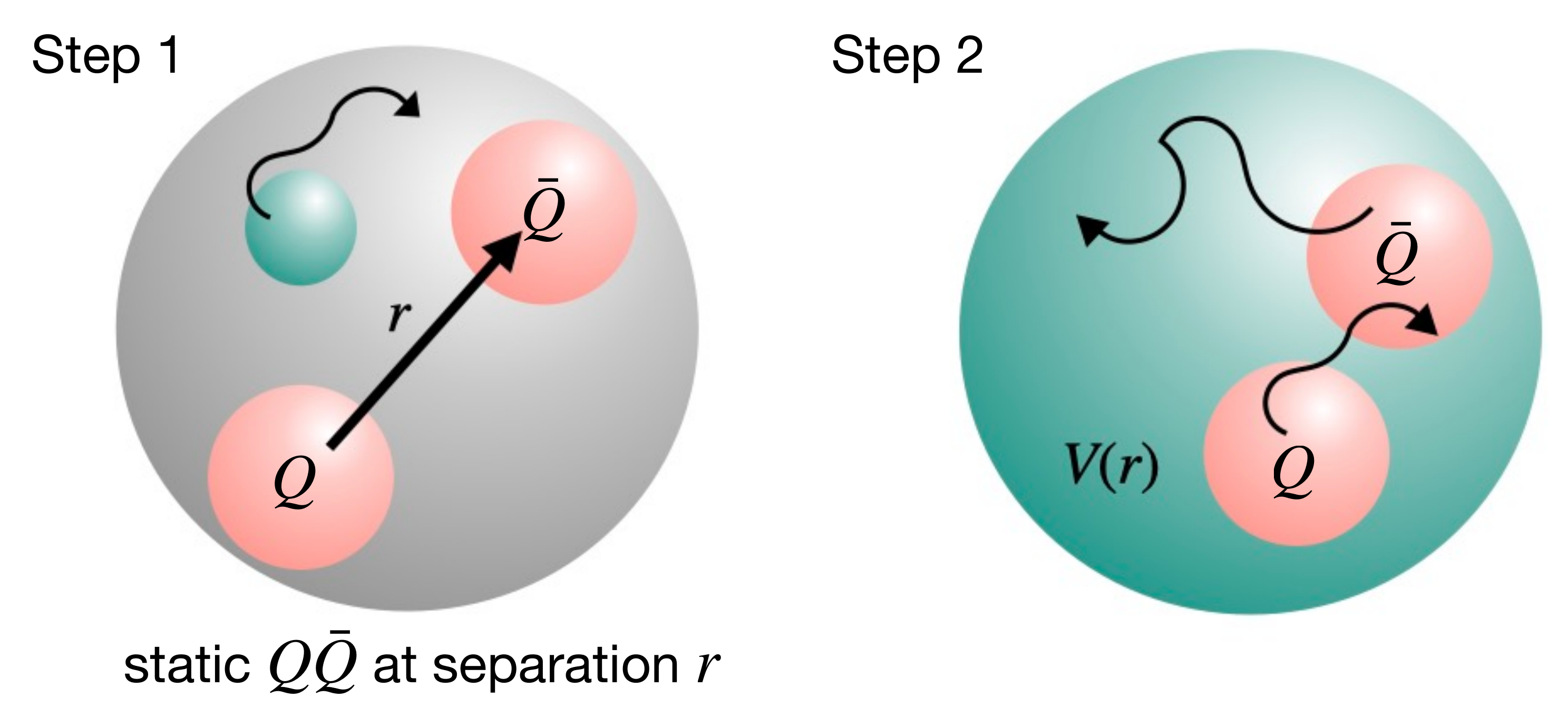}
    \caption{Sketch of the Born--Oppenheimer approach for heavy hybrid mesons.}
    \label{fig:bo-hybrid-schematic}
\end{figure}

A systematic approach to study conventional and hybrid quarkonium states is the Born--Oppen\-heimer effective field theory (BOEFT; see Refs.\ \cite{Berwein:2015vca,Brambilla:2017uyf,Soto:2020xpm,Berwein:2024ztx} and references therein). The Born--Oppen\-heimer approach consists of two steps. In step~1, the positions of the heavy quarks are fixed, and the light gluonic degrees of freedom are treated with lattice gauge theory, leading to quark-antiquark potentials $V(r)$. In step~2, the dynamics of the heavy quarks is studied using coupled-channel Schrödinger equations together with the potentials from step~1. The two steps are sketched in Figure~\ref{fig:bo-hybrid-schematic}.

The lattice gauge theory computation in step~1 provides the ordinary quarkonium static potential $V_{\Sigma_g^+}(r)$ and hybrid static potentials, most importantly the low-lying $V_{\Pi_u}(r)$ and $V_{\Sigma_u^-}(r)$ potentials (recent high precision lattice results for these potentials can be found in Ref.\ \cite{Schlosser:2021wnr}). Corrections due to the finite heavy quark mass $m_Q$ can be organized in powers of $1/m_Q$. At $\mathcal{O}(1/m_Q)$, there are four relevant potentials: spin-dependent interactions are described by $V_{11}^{sa}(r)$ and $V_{10}^{sb}(r)$ \cite{Brambilla:2018pyn,Brambilla:2019jfi,Soto:2023lbh}, while  $V_{\Sigma_u^-}^{\text{mix}}(r)$ and $V_{\Pi_u}^{\text{mix}}(r)$ couple hybrid and conventional quarkonium channels \cite{Oncala:2017hop}. 

The only existing lattice computation of these $1/m_Q$ potentials was exploratory, using only a single lattice spacing, gradient flow at a single flow time, and rather small volumes, where the latter restrict quark-antiquark separations to below about $0.5 \, \text{fm}$  \cite{Schlosser:2025tca}. Besides continuum and zero-flow time extrapolations, larger separations are particularly important, because heavy hybrid mesons are expected to have a size of around $1 \, \text{fm}$ or larger \cite{Capitani:2018rox}.

Determining the four potentials $V_{11}^{sa}(r)$, $V_{10}^{sb}(r)$, $V_{\Sigma_u^-}^{\text{mix}}(r)$ and $V_{\Pi_u}^{\text{mix}}(r)$ rigorously and precisely over a large range of quark-antiquark separations is the main objective of an ongoing long-term project, for which we provide a status report in this conference paper.

\section{Hybrid spin-dependent and hybrid-quarkonium
mixing potentials at $\mathcal{O}(1/m_Q)$}
\label{sec:bo-potentials}

The potentials we are interested in are related to matrix elements of the chromomagnetic operator \cite{Oncala:2017hop,Soto:2020xpm,Schlosser:2025tca}. Lattice results for these matrix elements diverge in the continuum limit and, thus, require renormalization. We regularize by applying gradient flow \cite{Luscher:2010iy} to our gauge link configurations, which introduces the flow time $ t_f$ as a regulator. At finite $t_f$, lattice results for the relevant matrix elements remain finite and can be converted, e.g.,\ to the $\overline{\text{MS}}$ scheme (see below).

Gradient flow smoothes the gauge field configurations over regions with radius $r_f = \sqrt{8t_f}$. Consequently, it suppresses the rather strong ultraviolet fluctuations introduced by the chromomagnetic operators, and, thereby, drastically reduces statistical errors.
However, the $SU(3)$ gauge theory corresponds to the limit $t_f \rightarrow 0$. To determine this limit numerically in a stable and controlled way, one should only consider data points for quark-antiquark separations $r$ where the flow time dependence is rather weak. In particular, this excludes separations $r \lesssim 2 r_f$ \cite{Schlosser:2025tca}.

At finite lattice spacing $a$ and flow time $t_f$, and when separating the static quark-antiquark pair along the $z$ axis, the four $\mathcal{O}(1/m_Q)$ hybrid spin-dependent and hybrid-quarkonium-mixing potentials are related to bare lattice matrix elements as follows:
\begin{eqnarray}
 & & V_{11}^{sa}(r,t_f,a)
=
i
\left\langle 0,\Pi_u^-\right|
g B_z(-r/2)
\left|0,\Pi_u^+\right\rangle (r)
\label{eq:nlo-matrix-element_1} \\
 & & V_{10}^{sb}(r,t_f,a)
=
i
\left\langle 0,\Sigma_u^-\right|
g B_y(-r/2)
\left|0,\Pi_u^+\right\rangle (r)
\label{eq:nlo-matrix-element_2} \\
 & &  m_Q V_{\Sigma_u^-}^{\text{mix}}(r,t_f,a)
=
\frac{i}{2}
\left\langle 0,\Sigma_g^+\right|
g B_z(-r/2)
\left|0,\Sigma_u^-\right\rangle (r) 
\label{eq:nlo-matrix-element_3} \\
 & & m_Q V_{\Pi_u}^{\text{mix}}(r,t_f,a)
=
\frac{i}{2}
\left\langle 0,\Sigma_g^+\right|
g B_x(-r/2)
\left|0,\Pi_u^+\right\rangle (r).
\label{eq:nlo-matrix-element_4}
\end{eqnarray}
The state $| 0, \Lambda_\eta^\epsilon \rangle$ denotes the lowest state containing a static quark at position $(0,0,+r/2)$ and a static antiquark at position $(0,0,-r/2)$ with quantum numbers $\Lambda_\eta^\epsilon$ (for a detailed discussion of these quantum numbers see e.g.\ Ref.\ \cite{Capitani:2018rox}).
The operator $g B_k(-r/2)$ with $k = x,y,z$ denotes the corresponding component of the chromomagnetic field at the position of the static antiquark and is discretized in this work using the standard clover leaf definition.
It is important to note that the validity of Eqs.\ (\ref{eq:nlo-matrix-element_1}) to (\ref{eq:nlo-matrix-element_4}) requires a specific definition of the phases of the states $|0,\Lambda_\eta^\epsilon\rangle$, as discussed in Ref.\ \cite{Schlosser:2025tca}.
 
The matrix elements appearing in Eqs.\ (\ref{eq:nlo-matrix-element_1}) to (\ref{eq:nlo-matrix-element_4}) and, consequently, the potentials $V_{11}^{sa}(r,t_f,a)$, $V_{10}^{sb}(r,t_f,a)$, $m_Q V_{\Sigma_u^-}^{\text{mix}}(r,t_f,a)$ and $m_Q V_{\Pi_u}^{\text{mix}}(r,t_f,a)$ can be extracted from suitably defined ratios of Wilson loops
\begin{equation}
\label{eq:generalized-ratio} R_{\Lambda_\eta^\epsilon\Lambda_{\eta'}^{\epsilon'}}^{B_k}(t;r,T)
=
\frac{W_{\Lambda_\eta^\epsilon\Lambda_{\eta'}^{\epsilon'}}^{B_k}(t;r,T)}
{\big(W_{\Lambda_\eta^\epsilon}(r,T)W_{\Lambda_{\eta}^{\epsilon'}}(r,T)\big)^{1/2}}
\left(
\frac{
W_{\Lambda_{\eta}^{\epsilon'}}(r,T/2-t)
W_{\Lambda_\eta^\epsilon}(r,T/2+t)}
{
W_{\Lambda_\eta^\epsilon}(r,T/2-t)
W_{\Lambda_{\eta}^{\epsilon'}}(r,T/2+t)}
\right)^{1/2}
\end{equation}
via
\begin{equation}
\label{EQN001} \lim_{T \rightarrow \infty} R_{\Lambda_\eta^\epsilon\Lambda_{\eta'}^{\epsilon'}}^{B_k}(t;r,T) = 
-i \big\langle 0,\Lambda_\eta^\epsilon\big|
g B_k(-r/2)
\big|0,\Lambda_{\eta}^{\epsilon'}\big\rangle(r)
.
\end{equation}
Here, $W_{\Lambda_\eta^\epsilon}(r,T)$ denotes a standard Wilson loop with spatial extent $r$, temporal extent $T$, and spatial transporters generating $\Lambda_\eta^\epsilon$ quantum numbers. The generalized Wilson loop $W_{\Lambda_\eta^\epsilon\Lambda_{\eta'}^{\epsilon'}}^{B_k}(t;r,T)$ contains an additional chromomagnetic field insertion on one of the temporal lines, shifted by $t$ relative to its midpoint.
Further details are given in Ref.\ \cite{Schlosser:2025tca}.

At finite flow time $t_f$, the continuum limit of the lattice potentials can be taken in a straightforward way, i.e.\
\begin{equation}
\label{EQN003} V_X(r,t_f) = \lim_{a \rightarrow 0} V_X(r,t_f,a) ,
\end{equation}
where $V_X = V_{11}^{sa} , V_{10}^{sb} , m_Q V_{\Sigma_u^-}^{\text{mix}} , m_Q V_{\Pi_u}^{\text{mix}}$. However, all four potentials $V_X(r,t_f)$ diverge in the $t_f \rightarrow 0$ limit. To compensate for these divergences, one has to multiply by a matching coefficient, which is known at next-to-leading order in the strong coupling for the $\overline{\text{MS}}$ scheme \cite{Brambilla:2023vwm,delaCruz:2024cix},
\begin{equation}
c_F(\mu,t_f) = 1-\frac{C_A \alpha_s(\mu)}{4\pi}
\ln\left(2\mu^2t_f e^{\gamma_E}\right)
+\mathcal{O}(\alpha_s^2).
\label{eq:flow-matching-factor}
\end{equation}
The renormalized potentials in the $\overline{\text{MS}}$ scheme are then given by
\begin{equation}
\label{EQN002} V_X(r,\mu) = \lim_{t_f \rightarrow 0} c_F(\mu,t_f) V_X(r,t_f) .
\end{equation}
To clarify our notation and its relation to the existing literature, the inverse of the coefficient (\ref{eq:flow-matching-factor}) was derived 
both in Ref.\ \cite{Brambilla:2023vwm}, where it was denoted by $c_F(t_f,\mu)$, and in Ref.\ \cite{delaCruz:2024cix}, where it was denoted by $Z(t_f,\mu)$, i.e., $c_F(\mu,t_f) = 1 / c_F(t_f,\mu) = 1 / Z(t_f,\mu)$.

The $\mu$ dependence in $V_X(r,\mu)$ cancels in the BOEFT against another perturbatively known matching coefficient
\cite{Eichten:1990vp},
\begin{equation}
    c_F^{(m_Q)}(\mu)=1+\frac{\alpha_s(\mu)}{2\pi}\bigg(C_F+C_A\bigg(1-\ln\bigg(\frac{m_Q}{\mu}\bigg)\bigg)\bigg)+\mathcal{O}(\alpha_s^2) .
    \label{eq:nrqcd-matching-factor}
\end{equation}
Finally, multiplication with this matching coefficient yields the four $\mathcal{O}(1/m_Q)$ potentials
\begin{equation}
V_X(r,m_Q) = c_F^{(m_Q)}(\mu) V_X(r,\mu)
\end{equation}
that enter the coupled-channel Schr\"odinger equations describing the states in the BOEFT, which are independent of $a$, $t_f$, and $\mu$.

\section{Lattice technicalities}
\label{sec:lattice}

\subsection{Gauge link configurations and interpolating operators}

For this work, we generated two new gauge link ensembles for pure $SU(3)$ gauge theory using the Wilson plaquette action and a heat-bath algorithm implemented in the CL2QCD software package \cite{Philipsen:2014mra}. The two ensembles have similar volumes in physical units, but different lattice spacings.
For both simulations, we used a cold start and thermalized with $20\,000$ heat-bath sweeps. Wilson loop averages were then computed on configurations separated by $10$ heat-bath sweeps at four different flow times, $t_f/a^2 = 0.2, 0.3, 0.4, 0.5$. See Table~\ref{tab:ensembles} for details.

\begin{table}[htb]
    \centering
    \small
    \begin{tabular}{ccccccc}
        \hline
        ensemble & $\beta$ & $a$ [fm] & $(L/a)^3 \times T/a$ & $L^3 \times T$
        & $r_f$ [fm] & \# conf. \\
        \hline
        A & $6.020$ & $0.0900$ & $48^4$
          & $(4.32\,\text{fm})^4$ & $0.114$, $0.139$, $0.161$, $0.180$
          & $3675$ \\
        B & $6.091$ & $0.0801$ & $54^4$
          & $(4.33\,\text{fm})^4$ 
          & $0.101$, $0.124$, $0.143$, $0.160$
          & $792$ \\
        \hline
    \end{tabular}
    \normalsize
    \caption{Gauge link ensembles. The lattice spacings in fm were obtained using Eq.\ (2.6) from Ref.\ \cite{Necco:2001xg}, setting $r_0 = 0.5 \, \text{fm}$. ``\# conf.'' denotes the number of gauge link configurations used for the computation of Wilson loop averages.}
    \label{tab:ensembles}
\end{table}

For the spatial transporters of the Wilson loops, we use the same interpolating operators as in Ref.\ \cite{Schlosser:2025tca}, which were optimized with respect to the generated ground state overlaps in Ref.\ \cite{Capitani:2018rox}. For additional ground-state enhancement, we use APE-smeared spatial links with $\alpha_{\text{APE}}=0.5$ and $N_{\text{APE}}=40$ and $N_{\text{APE}}=50$ smearing steps for ensembles A and B, respectively, where the parameters refer to the equations in Sec.\ 3.1.3 of Ref.\ \cite{Jansen:2008si}. 

\subsection{Extraction of potentials from ratios $R_{\Lambda_\eta^\epsilon\Lambda_{\eta'}^{\epsilon'}}^{B_k}(t;r,T)$}
\label{subsec:numerical-setup}

\begin{figure}[htb!]
    \centering
    \includegraphics[width=0.95\linewidth]
    {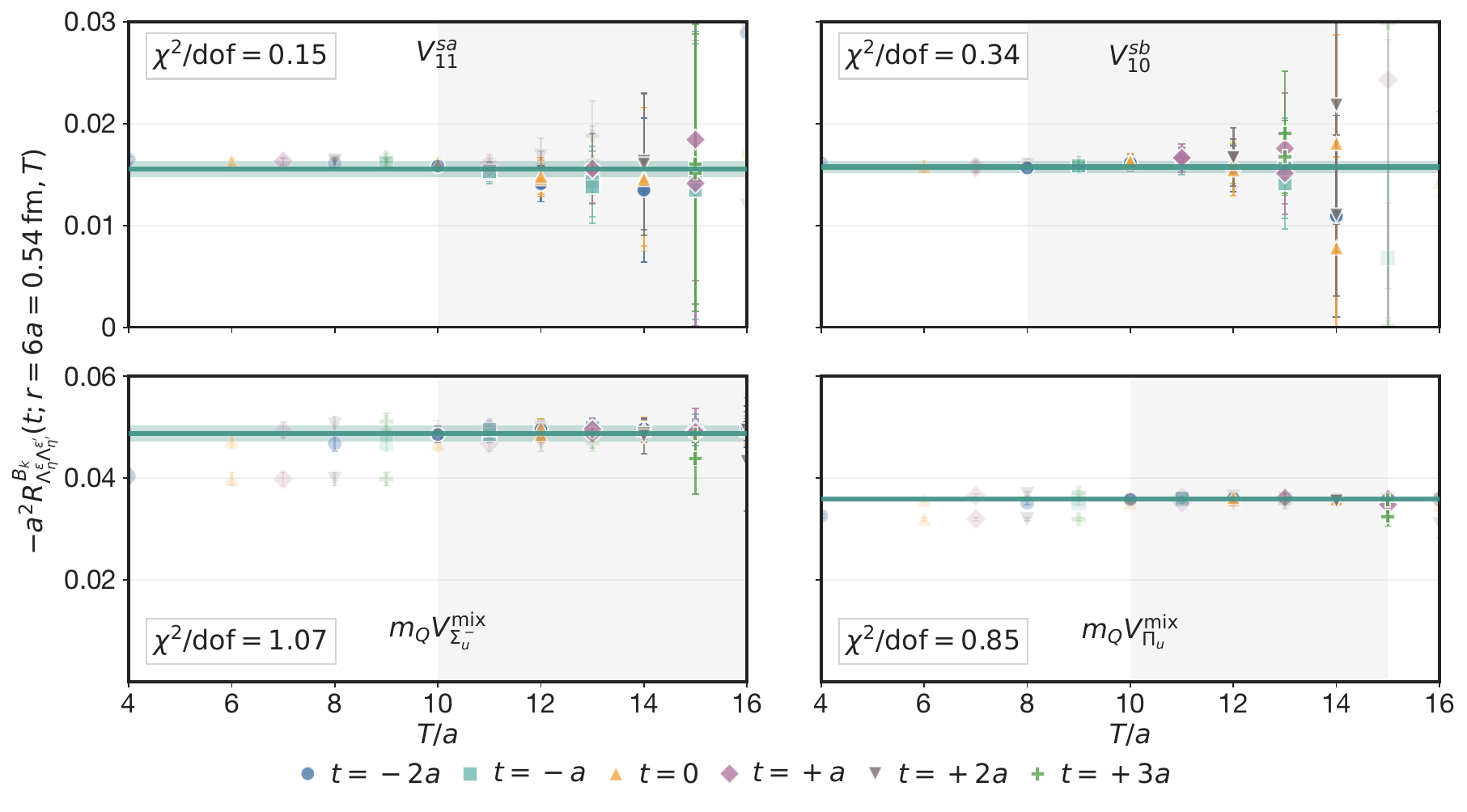}
    \caption{
    Representative ratios $R_{\Lambda_\eta^\epsilon\Lambda_{\eta'}^{\epsilon'}}^{B_k}(t;r,T)$ 
    and the corresponding fitted constants for the four $\mathcal{O}(1/m_Q)$ potentials on ensemble~A, at flow time $t_f/a^2=0.5$ (corresponding to $r_f=0.180\,\text{fm}$) and at quark-antiquark separation $r = 6 a = 0.54\,\text{fm}$.
    Colors indicate the shift $t$ of the chromomagnetic field insertion along one of the temporal lines relative to its midpoint.
    Only data points inside the shaded regions were included in the fits.}
    \label{fig:plateau-extraction}
\end{figure}

For both ensembles, the four considered flow times, 
each of the four potentials $V_{11}^{sa}(r,t_f,a)$, $V_{10}^{sb}(r,t_f,a)$, $m_Q V_{\Sigma_u^-}^{\text{mix}}(r,t_f,a)$ and $m_Q V_{\Pi_u}^{\text{mix}}(r,t_f,a)$ and all considered quark-antiquark separations, a potential data point was extracted from the corresponding ratio $R_{\Lambda_\eta^\epsilon\Lambda_{\eta'}^{\epsilon'}}^{B_k}(t;r,T)$ using Eqs.\ (\ref{eq:nlo-matrix-element_1}) to~(\ref{EQN001}). For each data point, we carried out an uncorrelated fit of a constant to the ratios in $(t, T)$ regions where excited states are strongly suppressed, i.e.\ the limit $T \rightarrow \infty$ in Eq.\ (\ref{EQN001}) is reached within statistical errors. We selected these ranges by following the strategy discussed in Ref.\ \cite{Schlosser:2025tca}, while at the same time keeping the ranges similar for all ensembles (A and B from this work and the ensemble used in Ref.\ \cite{Schlosser:2025tca}, whose data points we will compare against in the following). Examples of ratios and the corresponding fitted constants are shown for ensemble~A in Figure~\ref{fig:plateau-extraction} (here and in the following, statistical errors were determined with the jackknife method). The ratios are consistent with constants indicating that excited states are indeed negligible and that we are able to determine the $1/m_Q$ potentials reliably.

\section{Numerical results}
\label{sec:results}

To validate our lattice setup, we computed the ordinary static potential $V_{\Sigma_g^+}$ and the two lowest hybrid static potentials $V_{\Pi_u}$ and $V_{\Sigma_u^-}$ on ensemble A. Within the common range of separations and statistical uncertainties, our results agree with those of Ref.~\cite{Schlosser:2021wnr}. 
Since static potentials are not the focus of this contribution, we neither show nor discuss them here, but restrict the following discussion to the $\mathcal{O}(1/m_Q)$ potentials defined in Sec.\ \ref{sec:bo-potentials}.

\subsection{Lattice spacing and flow time dependence of the hybrid spin-dependent and hybrid-quarkonium mixing potentials}
\label{subsec:nlo-results}

\begin{figure}[htb]
    \centering
    \includegraphics[width=0.95\linewidth]
    {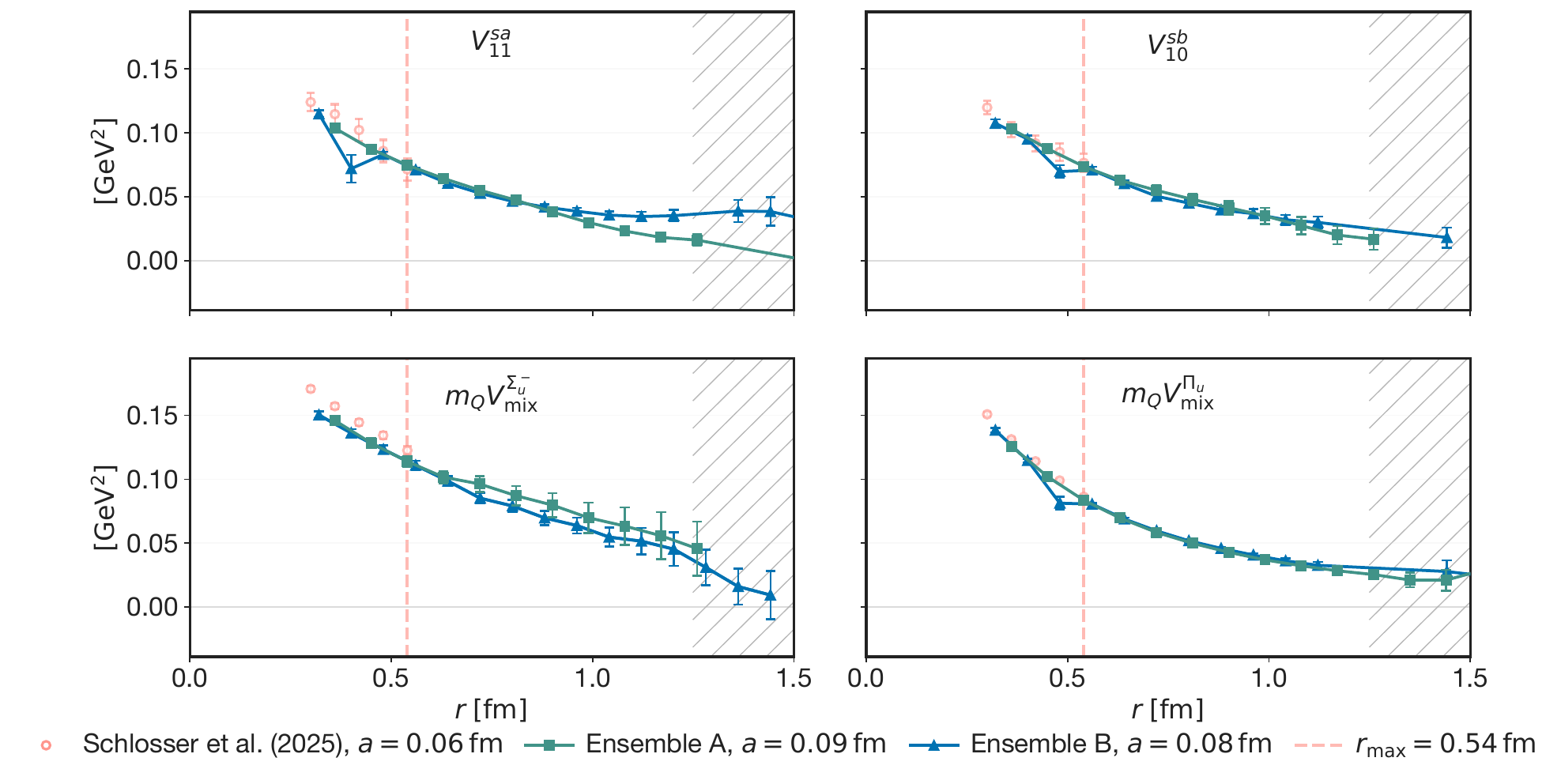}
    \caption{Comparison of the four $\mathcal{O}(1/m_Q)$ potentials on ensemble~A ($a = 0.0900 \, \text{fm}$, $r_f = 0.161 \, \text{fm}$), ensemble~B ($a = 0.0801 \, \text{fm}$, $r_f = 0.160 \, \text{fm}$)
    and a third ensemble used in Ref.~\cite{Schlosser:2025tca} ($a = 0.060 \, \text{fm}$, $r_f = 0.108 \, \text{fm}$). The maximum distance reached in the study~\cite{Schlosser:2025tca} is indicated by red dashed vertical lines.
    Data points inside the hatched regions are not stable with respect to variations of the $(t,T)$ regions (see Sec.\ \ref{subsec:numerical-setup}) and should be viewed with caution.
    }
    \label{fig:nlo-potentials-AB}
\end{figure}

We computed the hybrid spin-dependent potentials 
$V_{11}^{sa}(r,t_f,a)$ and $V_{10}^{sb}(r,t_f,a)$
and the hybrid-quarkonium mixing potentials $m_Q V_{\Sigma_u^-}^{\text{mix}}(r,t_f,a)$ and $m_Q V_{\Pi_u}^{\text{mix}}(r,t_f,a)$ on ensembles with different lattice spacings and for several flow times, with the aim of eventually performing continuum and zero-flow time extrapolations in accordance with Eqs.\ (\ref{EQN003}) and (\ref{EQN002}). We were able to reliably determine these potentials up to $r \approx 1.25\,\text{fm}$. Beyond this separation, the statistical precision rapidly deteriorates, and the extraction of potentials becomes strongly sensitive to the $(t,T)$ fitting ranges, resulting in significant systematic uncertainties, which are difficult to quantify. Results for ensemble~A ($a = 0.0900 \, \text{fm}$, $r_f = 0.161 \, \text{fm}$), ensemble~B ($a = 0.0801 \, \text{fm}$, $r_f = 0.160 \, \text{fm}$) and a third ensemble used in Ref.~\cite{Schlosser:2025tca} ($a = 0.060 \, \text{fm}$, $r_f = 0.108 \, \text{fm}$) are shown in Fig.\ \ref{fig:nlo-potentials-AB}.

\vspace{0.3cm}
\noindent \textbf{Lattice spacing dependence:} \\
For each of the four potentials, the data points from all three ensembles (corresponding to lattice spacings $a = 0.0900 \, \text{fm}$ [green points], $a = 0.0801 \, \text{fm}$ [blue points] and $a = 0.060 \, \text{fm}$ [red points]) are in fair agreement, within statistical uncertainties, with a single smooth curve. The lattice-spacing dependence therefore appears to be rather mild. Thus, we expect that we will be able to carry out a reliable and accurate continuum limit in the near future.

\vspace{0.3cm}
\noindent \textbf{Flow time dependence:} \\
The flow time dependence of the computed potentials is rather mild. For example at smaller separations $r \approx 0.6 \, \text{fm}$, where the dependence is stronger, after multiplying with their suppression factor $1/m_Q$, the differences between each of the four potentials at the smallest flow radius $r_f^\text{min} = 0.114 \, \text{fm}$ and at the largest flow radius $r_f^\text{max} = 0.180 \, \text{fm}$ (for ensemble~A) are of order $10 \, \text{MeV}$ for charm quarks, and reduce by a factor of 3 for bottom quarks. When performing the $t_f \rightarrow 0$ extrapolation according to Eq.\ (\ref{EQN002}), one has to multiply by the matching coefficient $c_F(\mu,t_f)$. This matching coefficient partly compensates for the flow time dependence, reducing the previously mentioned small differences even further by around 10\% to 50\%.
Thus, we expect that we will be able to carry out a reliable and accurate $t_f \rightarrow 0$ extrapolation in the near future.

\subsection{Impact of the calculated potentials on BOEFT predictions for hybrid mesons}
\label{subsec:BOeffect}

Lattice results for the hybrid spin-dependent and hybrid-quarkonium mixing potentials now extend to approximately $1.25\,\text{fm}$, more than doubling the previously available maximum separation $r_{\text{max}}=0.54\,\text{fm}$, indicated by the vertical red dashed line in Fig.\ \ref{fig:nlo-potentials-AB}. Although data points for separations larger than $0.54 \, \text{fm}$ are provided in the plots and tables of the exploratory study \cite{Schlosser:2025tca}, they should not be used in BOEFT predictions for hybrid mesons, because these separations
exceed half of the spatial lattice extent and thus might be affected by large systematic errors.

BOEFT predictions require the potentials over the full range of separations probed by the hybrid wave functions, which typically extend past $1\,\text{fm}$ \cite{Capitani:2018rox}. At sufficiently large separations, the potentials can reliably be described by the Effective String Theory~\cite{Luscher:2004ib}. The corresponding long-distance expressions, given in Eq.\ (14) of Ref.~\cite{Soto:2023lbh} for the spin-dependent potentials $V_{11}^{sa}$ and $V_{10}^{sb}$ and in Eq.~(40) of Ref.~\cite{Oncala:2017hop} for the mixing potential $V_{\Sigma_u^-}^{\text{mix}}$, all have a sign ambiguity ($V_{\Pi_u}^{\text{mix}}$ has none). With the previously available lattice data from Ref.\ \cite{Schlosser:2025tca} with maximum separation $0.54\,\text{fm}$, it was not possible to resolve these sign ambiguities \cite{TalkPaul}.
Our new lattice data, by extending to $1.25 \, \text{fm}$, come close to the region of validity of the Effective String Theory predictions and make nearly unambiguous predictions for the potentials $V_{11}^{sa}$, $V_{10}^{sb}$ and $V_{\Sigma_u^-}^{\text{mix}}$.

Consequently, our new results impact the
precision of BOEFT predictions positively, both for hybrid hyperfine splittings and spectra of conventional and hybrid quarkonia, by including their mixing.
The hybrid-quarkonium mixing potentials calculated in this work are not only relevant for heavy hybrid mesons, but also shed some light on the nature of conventional quarkonium states too.
In particular, a preliminary BOEFT calculation predicts a hybrid admixture of approximately $10\%$ for the $\eta_c(1S)$ state~\cite{TalksQCHS2026}.

\section{Outlook}
\label{sec:conclusions}

In this work, we presented hybrid spin-dependent and hybrid-quarkonium mixing potentials at finite lattice spacing and flow time.
The logical next step is to carry out continuum and zero-flow time extrapolations, which possibly require further simulations and computations. The computation of hybrid-quarkonium mixing potentials at short separations $r \ll 0.3 \, \text{fm}$ is another interesting future direction. Reaching such short separations requires fine lattices with $a \lesssim 0.04 \, \text{fm}$, where problems with topology freezing might arise \cite{Schaefer:2010hu,Schlosser:2021wnr}.
Moreover, sufficiently large physical volumes are required to accommodate the spatially extended hybrid wave functions, which implies a huge number of lattice sites and, thus, a significant amount of HPC resources.

\begin{acknowledgments}

We thank Carolin Schlosser for her exploratory work on the lattice determination of hybrid spin-dependent and hybrid-quarkonium mixing potentials,
which laid the foundation for the present work, as well as for sharing her contraction and analysis codes.
We thank Michael Eichberg for many helpful discussions,
in particular on spin corrections for conventional quarkonium and on technical aspects like gradient flow and matching.
We also acknowledge useful discussions with Nora Brambilla, Roberto Bruschini, Abhishek Mohapatra and Joan Soto.
We acknowledge funding by the Deutsche Forschungsgemeinschaft (DFG, German Research Foundation) -- project number 550503820.
The authors gratefully acknowledge the computing time provided to them at the NHR Center NHR@SW at Goethe University Frankfurt (project number 26459). This is funded by the Federal Ministry of Education and Research, and the state governments participating on the basis of the resolutions of the GWK for national high performance computing at universities (\texttt{www.nhr-verein.de/unsere-partner}).

\end{acknowledgments}


\begin{thebibliography}{99}

\bibitem{Berwein:2015vca}
M.~Berwein, N.~Brambilla, J.~Tarr{\'u}s Castell{\`a} and A.~Vairo,
\emph{Quarkonium hybrids with nonrelativistic effective field theories},
\href{https://doi.org/10.1103/PhysRevD.92.114019}
{Phys.\ Rev.\ D \textbf{92} (2015) no.~11, 114019}
[arXiv:1510.04299 [hep-ph]].

\bibitem{Brambilla:2017uyf}
N.~Brambilla, G.~Krein, J.~Tarr{\'u}s Castell{\`a} and A.~Vairo,
\emph{Born--Oppenheimer approximation in an effective field theory language},
\href{https://doi.org/10.1103/PhysRevD.97.016016}
{Phys.\ Rev.\ D \textbf{97} (2018) no.~1, 016016}
[arXiv:1707.09647 [hep-ph]].

\bibitem{Soto:2020xpm}
J.~Soto and J.~Tarr{\'u}s Castell{\`a},
\emph{Nonrelativistic effective field theory for heavy exotic hadrons},
\href{https://doi.org/10.1103/PhysRevD.102.014012}
{Phys.\ Rev.\ D \textbf{102} (2020) no.~1, 014012}
[erratum: Phys.\ Rev.\ D \textbf{110} (2024) no.~9, 099901]
[arXiv:2005.00552 [hep-ph]].

\bibitem{Berwein:2024ztx}
M.~Berwein, N.~Brambilla, A.~Mohapatra and A.~Vairo,
\emph{Hybrids, tetraquarks, pentaquarks, doubly heavy baryons, and quarkonia in Born--Oppenheimer effective theory},
\href{https://doi.org/10.1103/PhysRevD.110.094040}
{Phys.\ Rev.\ D \textbf{110} (2024) no.~9, 094040}
[arXiv:2408.04719 [hep-ph]].

\bibitem{Schlosser:2021wnr}
C.~Schlosser and M.~Wagner,
\emph{Hybrid static potentials in $SU(3)$ lattice gauge theory at small quark-antiquark separations},
\href{https://doi.org/10.1103/PhysRevD.105.054503}
{Phys.\ Rev.\ D \textbf{105} (2022) no.~5, 054503}
[arXiv:2111.00741 [hep-lat]].

\bibitem{Brambilla:2018pyn}
N.~Brambilla, W.~K.~Lai, J.~Segovia, J.~Tarr{\'u}s Castell{\`a} and A.~Vairo,
\emph{Spin structure of heavy-quark hybrids},
\href{https://doi.org/10.1103/PhysRevD.99.014017}
{Phys.\ Rev.\ D \textbf{99} (2019) no.~1, 014017}
[erratum: Phys.\ Rev.\ D \textbf{101} (2020) no.~9, 099902]
[arXiv:1805.07713 [hep-ph]].

\bibitem{Brambilla:2019jfi}
N.~Brambilla, W.~K.~Lai, J.~Segovia and J.~Tarr{\'u}s Castell{\`a},
\emph{QCD spin effects in the heavy hybrid potentials and spectra},
\href{https://doi.org/10.1103/PhysRevD.101.054040}
{Phys.\ Rev.\ D \textbf{101} (2020) no.~5, 054040}
[arXiv:1908.11699 [hep-ph]].

\bibitem{Soto:2023lbh}
J.~Soto and S.~T.~Valls,
\emph{Hyperfine splittings of heavy quarkonium hybrids},
\href{https://doi.org/10.1103/PhysRevD.108.014025}
{Phys.\ Rev.\ D \textbf{108} (2023) no.~1, 014025}
[arXiv:2302.01765 [hep-ph]].

\bibitem{Oncala:2017hop}
R.~Oncala and J.~Soto,
\emph{Heavy quarkonium hybrids: spectrum, decay and mixing},
\href{https://doi.org/10.1103/PhysRevD.96.014004}
{Phys.\ Rev.\ D \textbf{96} (2017) no.~1, 014004}
[arXiv:1702.03900 [hep-ph]].

\bibitem{Schlosser:2025tca}
C.~Schlosser and M.~Wagner,
\emph{Hybrid spin-dependent and hybrid-quarkonium mixing potentials at order $(1/m_Q)^1$ from $SU(3)$ lattice gauge theory},
\href{https://doi.org/10.1103/PhysRevD.111.074504}
{Phys.\ Rev.\ D \textbf{111} (2025) no.~7, 074504}
[arXiv:2501.08844 [hep-lat]].

\bibitem{Capitani:2018rox}
S.~Capitani, O.~Philipsen, C.~Reisinger, C.~Riehl and M.~Wagner,
\emph{Precision computation of hybrid static potentials in $SU(3)$ lattice gauge theory},
\href{https://doi.org/10.1103/PhysRevD.99.034502}
{Phys.\ Rev.\ D \textbf{99} (2019) no.~3, 034502}
[arXiv:1811.11046 [hep-lat]].

\bibitem{Luscher:2010iy}
M.~L{\"u}scher,
\emph{Properties and uses of the Wilson flow in lattice QCD},
\href{https://doi.org/10.1007/JHEP08(2010)071}
{JHEP \textbf{08} (2010), 071}
[erratum: JHEP \textbf{03} (2014), 092]
[arXiv:1006.4518 [hep-lat]].

\bibitem{Brambilla:2023vwm}
N.~Brambilla and X.~P.~Wang,
\emph{Off-lightcone Wilson-line operators in gradient flow},
\href{https://doi.org/10.1007/JHEP06(2024)210}
{JHEP \textbf{06} (2024), 210}
[arXiv:2312.05032 [hep-ph]].

\bibitem{delaCruz:2024cix}
D.~de la Cruz, A.~M.~Eller and G.~D.~Moore,
\emph{QCD field-strength correlators on a Polyakov loop with gradient flow at next-to-leading order},
\href{https://doi.org/10.1103/PhysRevD.110.094057}
{Phys.\ Rev.\ D \textbf{110} (2024) no.~9, 094057}
[arXiv:2410.01578 [hep-lat]].

\bibitem{Eichten:1990vp}
E.~Eichten and B.~R.~Hill,
\emph{Static Effective Field Theory: 1/m Corrections},
\href{https://doi.org/10.1016/0370-2693(90)91408-4}
{Phys.\ Lett.\ B \textbf{243} (1990), 427--431}.

\bibitem{Philipsen:2014mra}
O.~Philipsen, C.~Pinke, A.~Sciarra and M.~Bach,
\emph{CL$^2$QCD: lattice QCD based on OpenCL},
\href{https://doi.org/10.22323/1.214.0038}
{PoS \textbf{LATTICE2014} (2014), 038}
[arXiv:1411.5219 [hep-lat]].

\bibitem{Jansen:2008si}
K.~Jansen \textit{et al.} [ETM],
\emph{The static-light meson spectrum from twisted mass lattice QCD},
\href{https://doi.org/10.1088/1126-6708/2008/12/058}
{JHEP \textbf{12} (2008), 058}
[arXiv:0810.1843 [hep-lat]].

\bibitem{Necco:2001xg}
S.~Necco and R.~Sommer,
\emph{The $N_f=0$ heavy-quark potential from short to intermediate distances},
\href{https://doi.org/10.1016/S0550-3213(01)00582-X}
{Nucl.\ Phys.\ B \textbf{622} (2002), 328--346}
[arXiv:hep-lat/0108008 [hep-lat]].

\bibitem{Schaefer:2010hu}
S.~Schaefer \textit{et al.} [ALPHA],
\emph{Critical slowing down and error analysis in lattice QCD simulations},
\href{https://doi.org/10.1016/j.nuclphysb.2010.11.020}
{Nucl.\ Phys.\ B \textbf{845} (2011), 93--119}
[arXiv:1009.5228 [hep-lat]].

\bibitem{Luscher:2004ib}
M.~L\"uscher and P.~Weisz,
\emph{String excitation energies in SU(N) gauge theories beyond the free-string approximation},
\href{http://.doi.org/10.1088/1126-6708/2004/07/014}
{JHEP \textbf{07} (2004), 014}
[arXiv:hep-th/0406205 [hep-th]].

\bibitem{TalkPaul}
P.~H.~P\"utz,
\href{https://indico.cern.ch/event/1539475/contributions/6775627/attachments/3174843/5646431/Spin_dependent_potentials_for_hybrids.pdf}
{\emph{Spin dependent potentials for hybrids}},
talk presented at the Quarkonium Working Group Workshop 2025,
CERN, Geneva, 17--21 November 2025.

\bibitem{TalksQCHS2026}
V.~de~Jonge and P.~H.~P\"utz,
\href{https://indico.cern.ch/event/1531304/contributions/7106235/attachments/3306927/5917554/QCHS_Vilija_deJonge.pdf}
{\emph{Hybrid spin-dependent and hybrid-quarkonium mixing potentials at large quark-antiquark separations from lattice gauge theory}} and
\href{https://indico.cern.ch/event/1531304/contributions/7106232/attachments/3306975/5917048/QCHS_PaulPuetz_v2.pdf}
{\emph{The effect of spin and quarkonium-mixing potentials from lattice QCD on the heavy hybrid spectrum}},
talks presented at the XVII Conference on Quark Confinement and the Hadron Spectrum (QCHS 2026),
Wroc{\l}aw, Poland, 29 June--4 July 2026.

\end{thebibliography}
\end{document}